\documentclass[pra,twocolumn,showpacs,nobibnotes,floatfix,superscriptaddress]{revtex4}

\usepackage{latexsym,amssymb,amsfonts,mathbbol,graphicx,color}
\usepackage{dcolumn}
\usepackage{bm}

\begin{document}

\title{Quantum Error Correction Implementation after Multiple Gates} 
\author{Yaakov S. Weinstein}
\affiliation{Quantum Information Science Group, {\sc Mitre},
200 Forrestal Rd, Princeton, NJ 08540, USA}

\begin{abstract}
Correcting errors is a vital but expensive component of fault tolerant quantum computation. Standard fault tolerant protocol assumes the implementation of error correction, via syndrome measurements and possible recovery operations, after every quantum gate. In fact, this is not necessary. Here we demonstrate that error correction should be applied more sparingly. We simulate encoded single-qubit rotations within the [[7,1,3]] code and show via fidelity measures that applying error correction after every gate is not desirable. 
\end{abstract}


\maketitle

Quantum error correction (QEC) \cite{book,ShorQEC,CSS} is a necessary protocol for quantum computation but one that is very expensive in terms of number of qubits required and time to implement. Standard approaches to quantum fault tolerance (QFT), the computational framework that will allow for successful quantum computation despite a finite probability of error in basic computational gates \cite{Preskill,ShorQFT,G,AGP}, nevertheless assume that QEC is applied after every operation. In this paper we demonstrate that applying QEC after every operation is not necessary and in fact should not be done. This assertion is corroborated by simulating multiple single-logical-qubit operations on information encoded in the [[7,1,3]] QEC code \cite{Steane}. 

When implementing gates on encoded information we must ensure that the information does not leave the encoded space such that it may be subjected to errors. For many QEC codes universal quantum computation can be performed without leaving the encoded space if the gate set is restricted to Clifford gates plus the $T$-gate, a single-qubit $\pi/4$ phase rotation. A method for implementing an arbitrary single-qubit rotation (within prescribed accuracy $\epsilon$) with this restricted gate set was initially explored in \cite{SK1,SK2} and has recently become an area of intense investigation \cite{Svore1,KMM1,TMH,Svore2,KMM2,Selinger1,KMM3}. For Calderbank-Shor-Steane (CSS) codes, Clifford gates can be implemented bit-wise while the $T$-gates require a specially prepared ancilla state and a series of controlled-NOT gates. Thus, the primary goal of these investigations has been to construct circuits within $\epsilon$ of a desired (arbitrary) rotation while limiting the number of resource-heavy $T$-gates. As an example, $R_Z(.1)$ can be implemented with accuracy better than $10^{-5}$ using 78 \cite{Selinger1} or 56 \cite{KMM3} $T$-gates, interspersed by at least as many single-qubit Clifford gates.   
QFT would suggest that QEC be applied after each one of the more than 100 gates needed to implement such a rotation requiring thousands of additional qubits and hundreds of time steps. Thus, adhering to this tenet of QFT is very resource intensive. 

In line with recent work devoted to relaxing certain tenets of QFT while retaining reliability \cite{YSW,BHW,YSWTgate}, we demonstrate that QEC need not be applied after every gate and, in fact, should not be applied after every gate. Applying QEC less often will consume less resources, while still enabling successful quantum computation. This point was made, and addressed in a different way, in Ref.~\cite{WIPK}. Note that by application of QEC we refer to the implementation of syndrome measurements and possible recovery operations that must actively be applied during the computation. The entire computation, however, must be performed within a QEC encoding.  

The [[7,1,3]] QEC code will correct an error on one physical qubit of a seven qubit system that encodes one qubit of quantum information. If errors occur on two (physical) qubits the code will be unable to restore the system to its proper state. Let us assume a perfectly encoded state and (bit-wise implemented) Clifford gates that, with probability $p \ll 1$, (independently) cause an error on each qubit. The probability of an error on one qubit is then $7p-\mathcal{O}(p^2)$ and on two qubits $21p^2-\mathcal{O}(p^3)$. Thus, we can then be reasonably sure that at most only one qubit will have an error which will be corrected by QEC. Implementing two gates without applying QEC in between increases the probability that an error occurs to one qubit to $14p-\mathcal{O}(p^2)$ and that errors occur to two qubits to $84p^2-\mathcal{O}(p^3)$. The probability of errors on two qubits is still second order in $p$ and thus QEC applied after both gates will almost certainly correct the state of the system. When $n$ gates are applied the probability of an error on one qubit is $7n-\mathcal{O}(p^2)$ and on two qubits $21(n+2{n \choose 2})p^2$. Still, the probability that two (or more) errors occur remains of order $p^2$ and QEC will correct the single-qubit errors. Clearly QEC is not needed until the end of the gate sequence since at no point will the probability of errors on two or more qubits be of order $p$ (a similar argument can be made when including a two-qubit Clifford gate such as a controlled-NOT gate, this will be explored elsewhere). 

If $T$-gates are included the implementation becomes more complex. However, assuming the $T$-gate is done following the rules of QFT, two qubit errors will still occur with probability of order $p^2$ and $T$-gates will thus behave like the Clfford gates described above: QEC need not be applied until the end of the gate sequence. Of course, if QEC could be implemented perfectly, and we were not concerned with resource consumption, it would be worthwhile to apply QEC as much as possible. This will lower even further the possibility of multiple errors. However, QEC cannot be done perfectly in any realistic system (and resource usage is a concern). Thus, we are left to ask, how often should QEC be applied? Applying noisy QEC too often will be expensive in terms of time and qubits. Not applying QEC often enough will allow error probabilities to grow so large that errors become likely.  

To explore how often QEC should be applied we simulate single-qubit gates appropriate for the [[7,1,3]] QEC code in a nonequiprobable Pauli operator error environment \cite{QCC} with non-correlated errors. As in \cite{AP}, this model is a stochastic version of a biased noise model that can be formulated in terms of Hamiltonians coupling the system to an environment. Here, different error types arise with different arbitrary probabilities. Individual qubits undergo $\sigma_x^j$ errors with probability $p_x$, $\sigma_y^j$ errors with probability $p_y$, and $\sigma_z^j$ errors with probability $p_z$, where $\sigma_i^j$, $i = x,y,z$ are the Pauli spin operators on qubit $j$. We assume that only qubits taking part in a gate operation, initialization, or measurement will be subject to error while other qubits are perfectly stored. This idealized assumption is partially justified in that idle qubits may be less likely to undergo error than those involved in gates (see for example \cite{Svore}). In addition, we calculate accuracy measures only to second order in the error probabilities $p_i$ thus the effect of ignoring storage errors is likely minimal (this was observed in preliminary simulations). This latter point is buttressed by our above analysis demonstrating that two-qubit errors (for example) remain second order even after numerous gates and thus do not have a significant effect on the fidelity.
    
We start with an arbitrary single-qubit state, $|\psi\rangle=\cos\alpha|0\rangle+e^{i\beta}\sin\alpha|1\rangle$, perfectly encoded into the [[7,1,3]] error correction code. We then implement a series of gates, $...U_2U_1$, in the nonequiprobable error environment leading to a final state, $\rho_f$, of the 7 qubits. The final state is a function of the initial state, parameterized by $\alpha$ and $\beta$, and the error probabilities $p_x, p_y$, and $p_z$. We utilize two measures of accuracy comparing the simulated implementations with perfectly applied gates, $\rho_i$. The first is a state fidelity ${\rm{Tr}}[\rho_i\rho_f]$. The second is the logical gate fidelity, a state independent measure comparing the logical operation on the single-qubit of encoded information to the ideal single-qubit gate. To determine the logical gate fidelity we must first construct logical process matrices for the ideal and implemented operations. This is done by perfectly decoding $\rho_f$ and tracing over all qubits except the first giving the logical single-qubit output state. We then substitute $\alpha$ and $\beta$ for the specific states needed to calculate the process matrix \cite{QPT,book}. The logical gate fidelity is then simply ${\rm{Tr}}[\chi_i\chi_f]$ where $\chi_i$ is the process matrix of the perfect gate and $\chi_f$ is the process matrix of the implemented logical gate. 

After the gates, perfect (with no errors) or noisy (in the nonequiprobable error environment) QEC is applied to $\rho_f$ giving final states $\rho_{fp}$ and $\rho_{fn}$ respectively. Based on our above argument we expect perfect QEC to affirm the `correctability' of the errors that occur during implementation of multiple gates by raising the state or gate fidelity to unity (to at least second order in all $p_i$). In a realistic experiment, however, perfect QEC is not possible. Thus, we apply QEC in the nonequiprobable error environment to simulate a more realistic scenario. To apply noisy QEC in a fault tolerant fashion we utilize four qubit ancilla Shor states \cite{ShorQFT} for syndrome measurement. The Shor states are themselves constructed in the nonequiprobable error environment and construction is followed by one verification step \cite{WB}. Because every gate implemented in the nonequiprobable error environment has an error probability $p_i$ the fidelity of $\rho_{fn}$ will contain terms first order in $p_i$. Nevertheless, comparing $\rho_{fn}$ for single and multiple gates will alert us if there is a significant decrease in fidelity due to lack of error correction after every gate. 

We first look at gate sequences of only Clifford gates. Implementing a Clifford gate, $C$ on the [[7,1,3]] QEC code requires implementing $C^{\dag}$ on each of the 7 qubits. We choose sequences of Clifford gates typically found interspersed between $T$-gates in approximations of arbitary rotations: $H$, $PH$ and $HPH$, where $H$ is the Hadamard gate and $P = T^2$ is a $\pi/2$ phase gate \cite{Svore1,KMM1,Selinger1}. Results are shown in Table \ref{Cliff} up to first order in error probability (calculations were performed up to second order). Looking at both the state and logical gate fidelities for gate sequences with no error correction we see the expected decrease in fidelity as more gates are implemented. The decrease of the state fidelity is proportional to $7p_i$ the probability of single-qubit errors, as discussed above. 

Applying perfect error correction after one, two, or three Clifford gates gives state and logical gate fidelities of 1 (to third order). Applying noisy QEC after the sequence of gates we find that the state and gate fidelities are exactly the same for one and two Clifford gates. This demonstrates that there is no need to apply QEC after only one gate. Noisy QEC applied after three Clifford gates gives a lower fidelity state than when applied after two gates. In both the two and three gate case, noisy QEC causes a decrease in the state and gate fidelities when compared to the uncorrected state with respect to $\sigma_x$ errors and an increase with respect to $\sigma_z$ errors. Thus, if $\sigma_x$ errors are dominant one should implement even more gates before applying QEC. The decrease with respect to $\sigma_x$ errors can be attributed both to the fact that we have measured the bit-flip syndromes first (and thus uncorrected $\sigma_x$ errors occur during the phase-flip syndrome measurements), and to the use of noisy Shor states with one verification \cite{WB}. The fidelity due to $\sigma_y$ errors may increase or decrease upon application of noisy error correction.  

We compare these results to the case of applying QEC after each Clifford gate shown in the last line of Table~\ref{Cliff}. When a second gate is applied after the first application of QEC the fidelity decreases with respect to $p_x$ and $p_y$. The second application of QEC, however, increases the fidelity back to the same level as after the first QEC application. This implies that constant application of QEC will keep the fidelity steady. These simulations also underscore that there is no need to perform QEC after every gate as, after two gates, the gate and state fidelities are exactly the same whether or not QEC has been applied after the first gate.

\begingroup
\squeezetable
\begin{table*}
\caption{Fidelity measures of Clifford gates implemented in the nonequiprobable error environment with and without noisy error correction applied. We define $s_1 = \cos(4\alpha)$ and $s_2 = \cos(2\beta)\sin(2\alpha)^2$.}
\begin{tabular}{|c|c|c||c|c|c|}
\hline 
State Fidelity & no QEC & noisy QEC & Gate Fidelity & no QEC & noisy QEC \\\hline
H or P & $1-7p_x-7p_y-7p_z$ & $1-73p_x-19p_y-7p_z$ & H or P & $1-3p_x-5p_y-3p_z$ & $1-19p_x-5p_y-3p_z$ \\\hline
PH  & $1-14p_x-14p_y-14p_z$ & $1-73p_x-19p_y-7p_z$ & PH & $1-8p_x-8p_y-6p_z$ & $1-19p_x-5p_y-3p_z$ \\\hline
HPH  & $1-21p_x-21p_y-21p_z$ & $1-\frac{1}{2}(155-s_1-2s_2)p_x$ & HPH & $1-11p_x-13p_y-9p_z$ & $1-23p_x-11p_y-8p_z$\\
 & & $-\frac{1}{4}(97-3s_1-6s_2)p_y-\frac{1}{4}(61-3s_1-6s_2)p_z$ & & & \\\hline
P-QEC-H & $1-80p_x-26p_y-14p_z$ & $1-73p_x-19p_y-7p_z$ & P-QEC-H & $1-54p_x-15p_y-6p_z$ & $1-19p_x-5p_y-3p_z$ \\\hline
\end{tabular}
\label{Cliff}
\end{table*}
\endgroup

We now look at sequences of gates that include a $T$-gate. To implement a logical $T$-gate on a state encoded in the [[7,1,3]] QEC code requires constructing the ancilla state $|\Theta\rangle = \frac{1}{\sqrt{2}}(|0_L\rangle+e^{i\frac{\pi}{4}}|1_L\rangle)$, where $|0_L\rangle$ and $|1_L\rangle$ are the logical basis states on the [[7,1,3]] QEC code. Bit-wise CNOT gates are then applied between the state $|\Theta\rangle$ and the encoded state with the $|\Theta\rangle$ state qubits as control. Measurement of zero on the encoded state projects the encoded state with the application of a $T$-gate onto the qubits that had made up the $|\Theta\rangle$ state. 

\begin{figure}
\includegraphics[width=8.5cm]{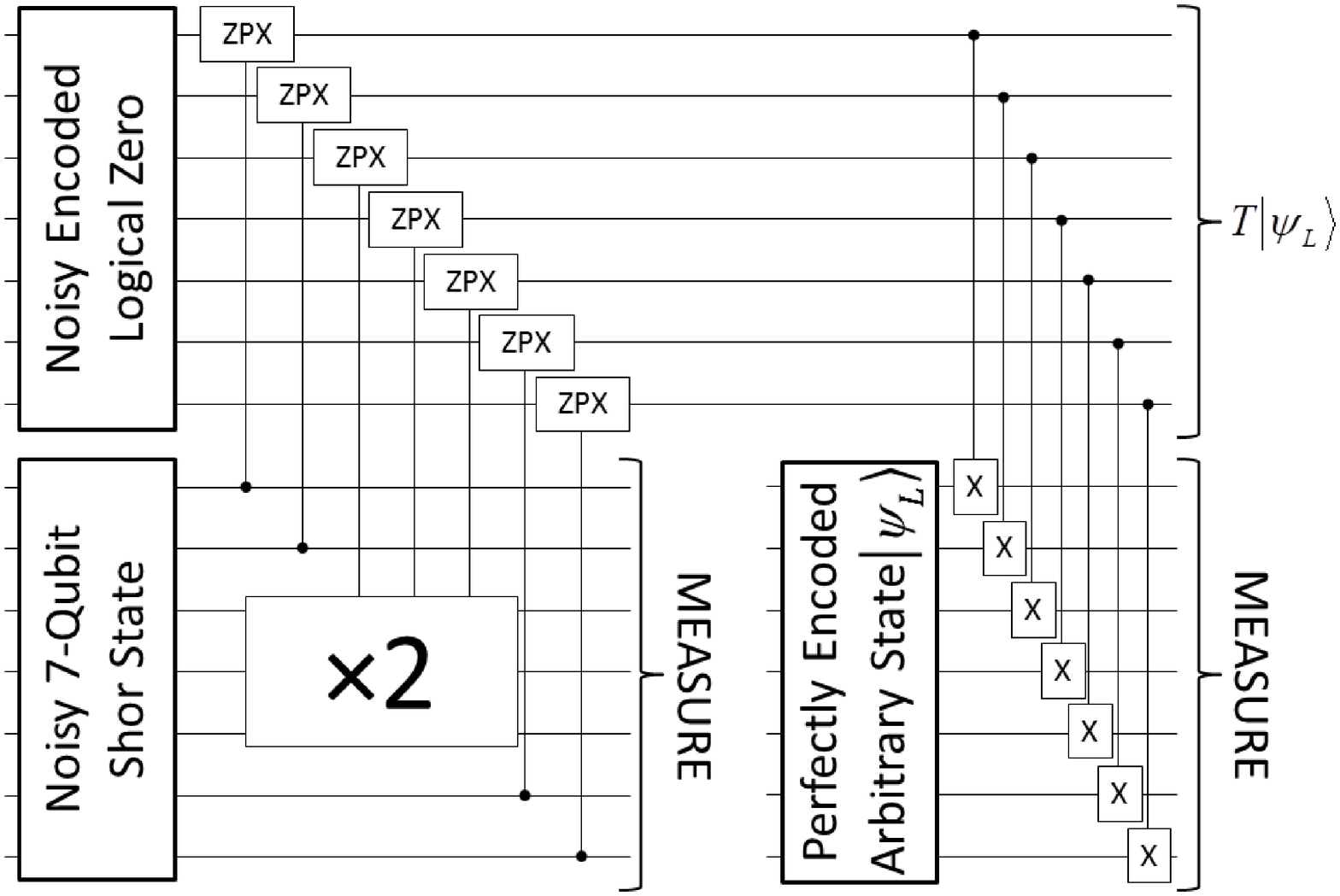}
\caption{Implementation of [[7,1,3]] QEC code $T$-gate. }
\label{Tgate}
\end{figure}

\begin{figure}
\includegraphics[width=8.5cm]{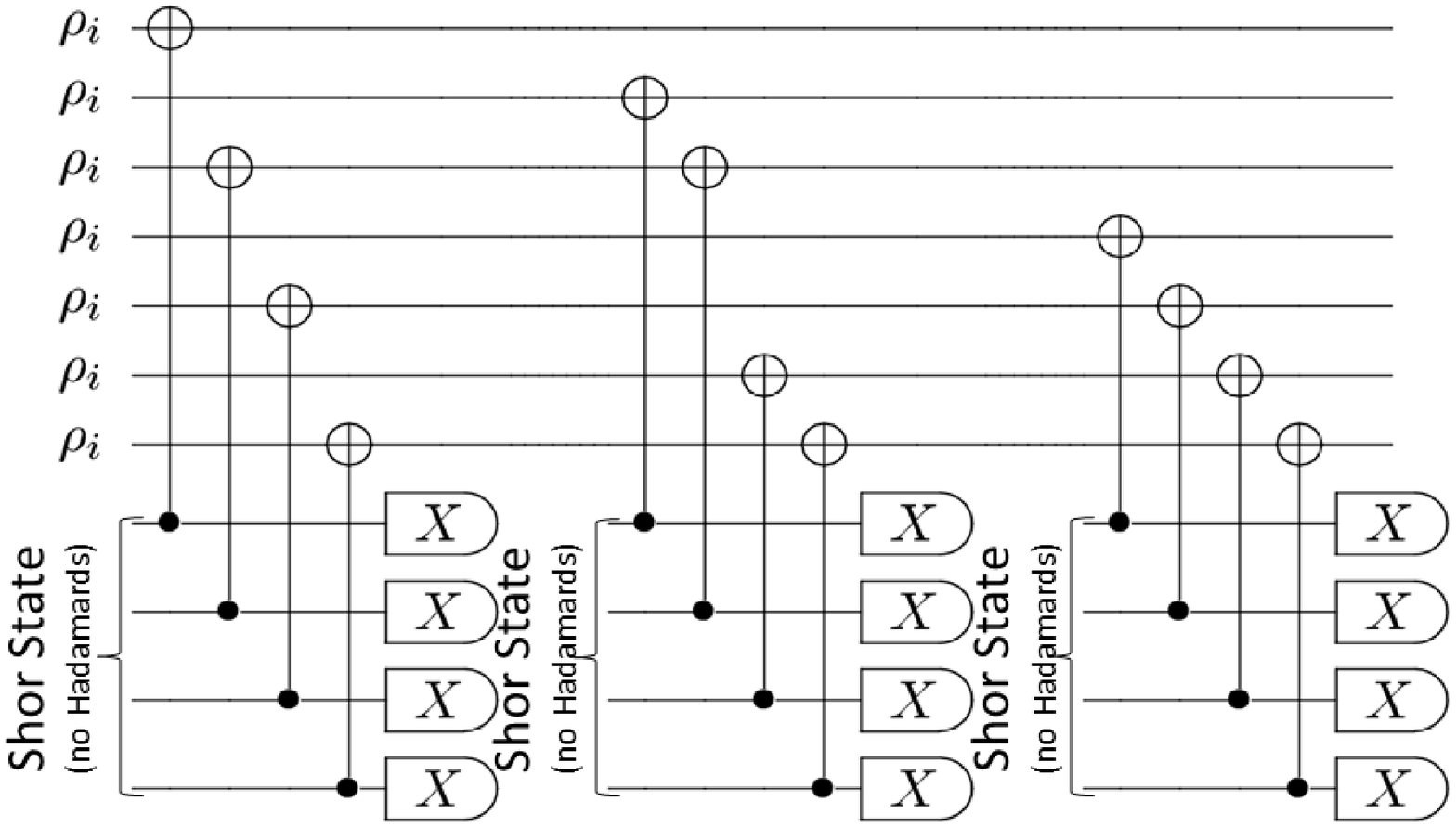}
\includegraphics[width=5cm]{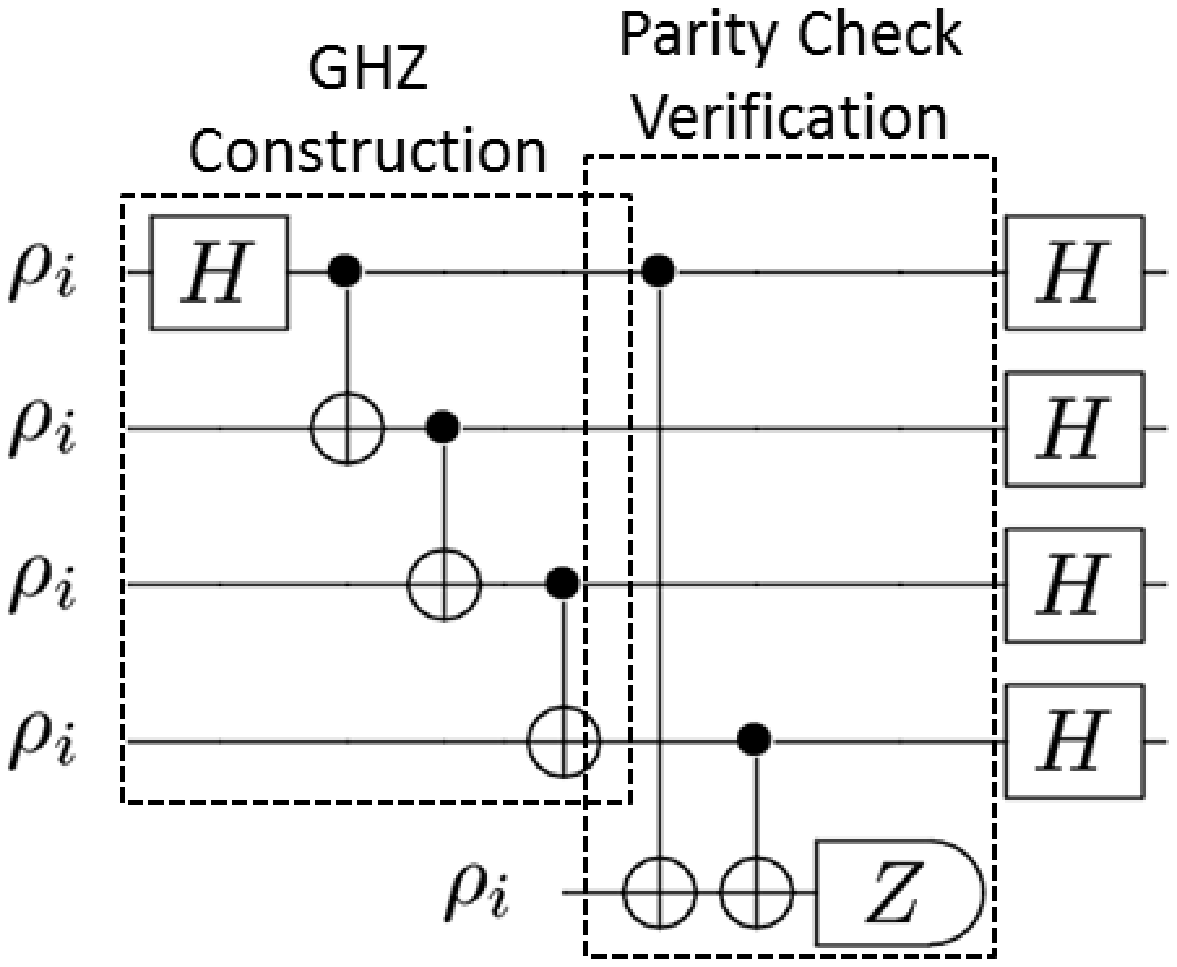}
\caption{Top: Circuit for phase syndrome measurement on the [[7,1,3]] QEC code, used here to initialize logical zero state.  
Bottom: Four qubit Shor state construction with one verification.}
\label{QEC}
\end{figure}

To ensure fault tolerance in the construction of $|\Theta\rangle$ requires the following: (1) A logical zero state is encoded by applying error correction to 7 qubits all initially in the state $|0\rangle$ \cite{Preskill}. We use Shor state ancilla for syndrome measurements \cite{ShorQFT}. (2) A seven qubit Shor state \cite{ShorQFT} in constructed and proper verifications are applied. (3) Seven controlled-ZPX gates: 
\begin{eqnarray}
C(ZPX) &=& 
\left( 
\begin{array}{cccc}
1 & 0 & 0 & 0 \\
0 & 1 & 0 & 0 \\
0 & 0 & 0 & e^{i\frac{\pi}{4}} \\
0 & 0 & e^{-i\frac{\pi}{4}} & 0 \\
\end{array}
\right),
\end{eqnarray}
are applied each between a qubit of the Shor state and a qubit of the logical zero state with the Shor state qubits as control. (4) Measurement of the Shor state (with even parity outcome) completes the projection and the construction of the logical state $|\Theta\rangle$. Circuits for these steps are shown in Figs.~\ref{Tgate} and \ref{QEC}. Our simulations, done in the nonequiprobable error environment, follow the implementation method of \cite{YSWTgate} and, after application of perfect QEC, are of unit fidelity.

In Table \ref{T} we compare the simulation of the $T$-gate alone with that of a $T$-gate with one, two, or three Clifford gates. Once again, the question we are trying to address is how often noisy QEC should be applied. 

\begingroup
\squeezetable
\begin{table*}
\caption{Fidelity measures of Clifford gates implemented in the nonequiprobable error environment with and without noisy error correction applied. }
\begin{tabular}{|c|c|c||c|c|c|}
\hline 
State Fidelity & no QEC & noisy QEC & Gate Fidelity & no QEC & noisy QEC \\\hline
T & $1-7p_x-7p_y-26p_z$ & $1-73p_x-19p_y-7p_z$ & T & $1-3p_x-5p_y-14p_z$ & $1-19p_x-5p_y-3p_z$ \\\hline
PT  & $1-14p_x-14p_y-33p_z$ & $1-73p_x-19p_y-7p_z$ & PT & $1-8p_x-8p_y-17p_z$ & $1-19p_x-5p_y-3p_z$ \\\hline
HT  & $1-14p_x-14p_y-33p_z$ & $1-73p_x-19p_y-7p_z$ & HT & $1-6p_x-10p_y-17p_z$ & $1-19p_x-5p_y-3p_z$ \\\hline
TPH  & $1-7p_x-7p_y-40p_z$ & $1-73p_x-19p_y-7p_z$ & TPH & $1-3p_x-5p_y-20p_z$ & $1-19p_x-5p_y-3p_z$ \\\hline
THPH & $1-14p_x-14p_y-33p_z$ & $1-73p_x-19p_y-7p_z$ & THPH & $1-6p_x-8p_y-17p_z$ & $1-19p_x-5p_y-3p_z$ \\\hline
P-QEC-T & $1-73p_x-19p_y-7p_z$ & $1-73p_x-19p_y-7p_z$ & P-QEC-T & $1-19p_x-5p_y-3p_z$ & $1-19p_x-5p_y-3p_z$ \\\hline
\end{tabular}
\label{T}
\end{table*}
\endgroup

The first point of interest is the difference in fidelities between the $T$-gate and the Clifford gates $P$ and $H$. The fidelity of these gates as a function of $p_x$ and $p_y$ is the same. The $T$-gate, however, is much more sensitivite to $\sigma_z$ errors. Thus, the accuracy `cost' (there is, of course, a prohibitive cost in the number of extra qubits utilized and the time of implementation) of applying a $T$-gate as opposed to a single-qubit Clifford gate is only with respect to phase errors. Applying noisy QEC to the single gates equalizes the fidelities of the $T$-gate and Clifford gates. Presumably, this is because the first order error terms arising from the implementation of the gate are corrected by QEC and the remaining first order error terms are due to the QEC itself.  

Implementing a single Clifford gate after a $T$-gate decreases the fidelity by the same amount as applying a Clifford gate after another Clifford gate. Implementing a $T$-gate after two Clifford gates decreases the fidelity compared to the $T$-gate alone only with respect to $p_z$ and, in fact, the fidelity with respect to $p_x$ and $p_y$ is higher than that of two Clifford gates alone. Implementing the $T$ gate after three Clifford gates gives fidelity equal to applying a Clifford gate after the $T$ gate. Thus, the simulations show some complexity in terms of which gates will decrease or increase the fidelity with respect to the different error types. 

Applying perfect QEC after any of the above gate sequences gives unit fidelity to second order in $p_i$ (as opposed to third order for sequences of only Clifford gates). As expected, the errors are correctable. 

We noted above that the fidelity measures after noisy QEC appear to be insensitive to the gates applied before QEC. This is clearly seen in Table \ref{T}. Presumably this arises because the QEC corrects the errors of the previous gates or at least increases their order in error probability (in line with the perfect error correction simulations), and the noise inherent in the QEC is solely responsible for the first order error terms. Note that, unlike the case of all Clifford gates, when three or four gates are applied before QEC with one being a $T$-gate the fidelity of the state after QEC remains constant. This implies that one can apply three of four gates without QEC.

These simulations should also be compared to the case of applying QEC after an initial $T$-gate and then again after a Clifford gate. The results of this latter simulation with the $P$ gate are shown in the last line of Table~\ref{T}. Implementation of $P$ after the initial QEC does not change the fidelity and the fidelity remains constant upon the second application of QEC. This implies that constant application of QEC will keep the fidelity steady and again underscores that there is no need to perform QEC after every gate. 

In conclusion, we have explored the question of how often quantum error correction needs to applied during a sequence of logical single-qubit gates from the gate set Clifford plus $T$ as would be necessary for the implementation of arbitrary single-qubit rotations. Our analysis demonstrated that QEC is actually necessary only at the end of such a sequence as two-qubit errors do not occur to first order in error probability. In addition, simulated implementations in which QEC is imperfect demonstrate that for a sequence of Clifford gates it appears useful to apply QEC not more often than every other gate. However, if the sequence includes a $T$-gate there is no need to apply error correction before at least four gates. All of our simulations were done within the [[7,1,3]] QEC code but the results should be directly applicable to other CSS codes and, perhaps, to other QEC codes as well. 

In addition, we have utilized logical $\chi$-matrices in evaluating the logical gate fidelity of the Clifford and $T$-gates. These may prove generally useful in simulating quantum fault tolerance. Finally, we note that the application of noisy QEC should be taken into account when deciding to what accuracy arbitrary rotations should be implemented. 

I would like to thank G. Gilbert for insightful comments and S. Buchbinder for help on the initial stages of this work. This research is supported under MITRE Innovation Program Grant 51MSR662.

\end{document}